\documentclass[final]{agujournal2019}
\usepackage{url} %this package should fix any errors with URLs in refs.
\usepackage{lineno}
\usepackage[inline]{trackchanges} %for better track changes. finalnew option will compile document with changes incorporated.
\usepackage{soul}
\draftfalse

\journalname{Geophysical Research Letters}

\begin{document}

%% ------------------------------------------------------------------------ %%
%  Title
%
% (A title should be specific, informative, and brief. Use
% abbreviations only if they are defined in the abstract. Titles that
% start with general keywords then specific terms are optimized in
% searches)
%
%% ------------------------------------------------------------------------ %%

% Example: \title{This is a test title}

\title{Detecting local earthquakes via fiber-optic cables in telecommunication conduits under Stanford University campus using deep learning}

%% ------------------------------------------------------------------------ %%
%
%  AUTHORS AND AFFILIATIONS
%
%% ------------------------------------------------------------------------ %%

% Authors are individuals who have significantly contributed to the
% research and preparation of the article. Group authors are allowed, if
% each author in the group is separately identified in an appendix.)

% List authors by first name or initial followed by last name and
% separated by commas. Use \affil{} to number affiliations, and
% \thanks{} for author notes.
% Additional author notes should be indicated with \thanks{} (for
% example, for current addresses).

% Example: \authors{A. B. Author\affil{1}\thanks{Current address, Antartica}, B. C. Author\affil{2,3}, and D. E.
% Author\affil{3,4}\thanks{Also funded by Monsanto.}}

\authors{Fantine Huot\affil{1}, Robert Clapp\affil{1}, and Biondo L. Biondi\affil{1}}

\affiliation{1}{Stanford University}
% \affiliation{2}{Second Affiliation}
% \affiliation{3}{Third Affiliation}
% \affiliation{4}{Fourth Affiliation}

\affiliation{1}{397 Panama Mall, University Stanford, CA 94305, USA}
%(repeat as many times as is necessary)

%% Corresponding Author:
% Corresponding author mailing address and e-mail address:

% (include name and email addresses of the corresponding author.  More
% than one corresponding author is allowed in this LaTeX file and for
% publication; but only one corresponding author is allowed in our
% editorial system.)

% Example: \correspondingauthor{First and Last Name}{email@address.edu}

\correspondingauthor{Fantine Huot}{fantine@sep.stanford.edu}

%% Keypoints, final entry on title page.

%  List up to three key points (at least one is required)
%  Key Points summarize the main points and conclusions of the article
%  Each must be 140 characters or fewer with no special characters or punctuation and must be complete sentences

% Example:
% \begin{keypoints}
% \item	List up to three key points (at least one is required)
% \item	Key Points summarize the main points and conclusions of the article
% \item	Each must be 140 characters or fewer with no special characters or punctuation and must be complete sentences
% \end{keypoints}

\begin{keypoints}
\item We use fiber-optic cables in telecommunication conduits under Stanford University campus to record seismicity

\item We train a neural network to automatically detect small local earthquakes on the recorded data

\item We detect small uncataloged local earthquakes and demonstrate that our fiber-optic acquisition can complement sparse seismometer networks
\end{keypoints}

%% ------------------------------------------------------------------------ %%
%
%  ABSTRACT and PLAIN LANGUAGE SUMMARY
%
% A good Abstract will begin with a short description of the problem
% being addressed, briefly describe the new data or analyses, then
% briefly states the main conclusion(s) and how they are supported and
% uncertainties.

% The Plain Language Summary should be written for a broad audience,
% including journalists and the science-interested public, that will not have 
% a background in your field.
%
% A Plain Language Summary is required in GRL, JGR: Planets, JGR: Biogeosciences,
% JGR: Oceans, G-Cubed, Reviews of Geophysics, and JAMES.
% see http://sharingscience.agu.org/creating-plain-language-summary/)
%
%% ------------------------------------------------------------------------ %%

%% \begin{abstract} starts the second page

\begin{abstract}
With fiber-optic seismic acquisition development, continuous dense seismic monitoring is becoming increasingly more accessible. Repurposing fiber cables in telecommunication conduits makes it possible to run seismic studies at low cost, even in locations where traditional seismometers are not easily installed, such as in urban areas. However, due to the large volume of continuous streaming data, data collected in such a manner will go to waste unless we significantly automate the processing workflow. We train a convolutional neural network (CNN) for earthquake detection using data acquired over three years by fiber cables in telecommunication conduits under Stanford University campus. We demonstrate that fiber-optic systems can effectively complement sparse seismometer networks to detect local earthquakes. The CNN allows for reliable earthquake detection despite a low signal-to-noise ratio and even detects small-amplitude previously-uncataloged events.
\end{abstract}

\section*{Plain Language Summary}
New technological developments allow us to record vibrations inside the Earth using fiber-optic cables, the same type of cables that constitute the backbone of the internet. We use the fiber cables in the telecommunication conduits under Stanford University campus to record vibrations over 3 years and monitor local earthquakes. This system records large data volumes that require automated processing methods. We train a machine learning model for earthquake detection on these data. This machine learning model helps us identify previously undetected small earthquakes. 

\section*{Keywords}
urban geophysics, distributed acoustic sensing, DAS, acquisition, earthquake, machine learning

\section{Introduction}

Local small-amplitude earthquakes can provide valuable information about an area's fault location and characteristics \cite{stoffer2006s,field2015ucerf3}. A better understanding of local fault structures helps minimize risk in earthquake-prone areas, especially in urban areas with high population density. Cities are sometimes equipped with strong-motion sensors to monitor medium to large magnitude earthquakes. Still, these sensors do not record small local events \cite{lay2002global} that could give additional clues as to precise fault location. Recording small-amplitude earthquakes requires a dense array of continuously recording seismometers. Such sensors, however, are rarely deployed in the middle of cities, where the anthropogenic seismic noise is high, and sensors are prone to tampering. Moreover, seismometers can only provide sparse geographical coverage and are thus not necessarily close to the seismic events of interest.

Over the last decade, a technology called distributed acoustic sensing (DAS) has been developed that uses fiber-optic cables to record strain along the cable. 
It is increasingly adopted in the energy industry, with applications ranging from %microseismic monitoring \cite{webster2013micro}, 
hydraulic fracturing monitoring \cite{bakku2015fracture}, CO$_2$ sequestration monitoring \cite{daley2013field}, geothermal monitoring \cite{lellouch2020comparison,lellouch2021low}, to time-lapse imaging of reservoirs \cite{mateeva2013distributed}.
%,miller2016simultaneous}.
More recently, DAS acquisitions are deployed in the near-surface for earthquake monitoring \cite{lindsey2017fiber,biondi2021using}. Repurposing the fiber cables from the existing telecommunication infrastructure makes it possible to record dense continuous seismic data in urban areas at low cost. The additional spatial information provided by dense recordings is valuable for identifying, characterizing, and separating signal from noise. From 2016 to 2019, we connected a DAS interrogator unit to the fiber cables in telecommunication conduits under Stanford University campus, recording years of continuous seismic data \cite{huot2017automatic,martin2018seismic,martin2018eighteen,biondi2021scaling}. 

Taking advantage of pre-existing fibers for seismic acquisition shows excellent promise for earthquake monitoring. In 2019, a rapid-response experiment was deployed using the infrastructure  fiber cables along nearby highways after the M 7.1 Ridgecrest earthquake in California \cite{karrenbach2019rapid,li2021rapid}. By applying template matching to the recorded data, they detected abundant aftershocks on multiple faults near the epicenter of the mainshock. Given the widespread fiber networks worldwide, DAS can deliver fast and high-resolution seismic monitoring and promote better understanding of earthquake physics.

The processing of dense continuous acquisitions, however, can quickly become computationally prohibitive due to the large amounts of recorded data. The Stanford DAS array, for instance, recorded continuously %600 channels
at 50 samples per second for three years, requiring automated processing methods to uncover the data's full potential. Furthermore, the signal-to-noise ratio (SNR) of the recordings can vary significantly along the cable. The coupling with the ground varies at different portions of the conduits, and anthropogenic sources such as car traffic generate strong amplitude noise. These limitations render this type of data unsuitable for conventional thresholding-based earthquake detection methods such as  short-term average to long-term average (STA/LTA).
 
In the Ridgecrest study mentioned above, the aftershocks were detected by template matching, which requires an initial waveform template. Such a template is not readily available for the detections of small events on undocumented faults. An attractive alternative to this approach is the usage of supervised learning. Once a machine learning (ML) model is trained, application to newly acquired fiber data, possibly even continuously, can potentially remove many bottlenecks of traditional processing. Neural networks have been shown to be successful at earthquake detection on seismometer recordings \cite{perol2018convolutional,huot2018jump,zhang2019aftershock,zheng2020sc}. Recent studies demonstrate that they also have great potential for waveform-based event detection on continuous fiber recordings in the presence of coherent noise \cite{huot2017automatic,martin2018seismic,huot2018automated,huot2018machine}.

Herein, we train a convolutional neural network (CNN) for earthquake detection on the three years of data recorded by the Stanford DAS array. Once the network is trained, it can process continuous data in a matter of seconds, allowing real-time processing to be done using only one V100 GPU. This study is a prototype experiment for repurposing infrastructure fiber cables for seismic analysis. The acquisition presented here is but a first attempt and has several limitations. First, the coupling with the ground is not very good and varies significantly over different portions of the fiber, affecting the signal quality throughout the recorded data. The recording system was not the latest generation, and systems available today would deliver significant improvements in the SNR of the recorded data. In addition, this array is relatively small, 600 m along its widest dimension, limiting precise event localization and source mechanism analysis. Indeed, source localization involves solving an inverse problem, matching arrival times at different locations. A small array does not provide sufficient aperture to achieve useful resolution. Therefore, this study focuses primarily on the array's detection capabilities. We demonstrate that despite these limitations, this DAS array can detect small local earthquakes and effectively complement the sparse seismometer network in the area. Larger DAS arrays should provide the information needed for the event localization.

\section{The Stanford DAS array} 

\begin{figure}
    \begin{center}
        \includegraphics[width=0.45\linewidth,trim={3.5cm 0cm 3cm 1cm},clip]{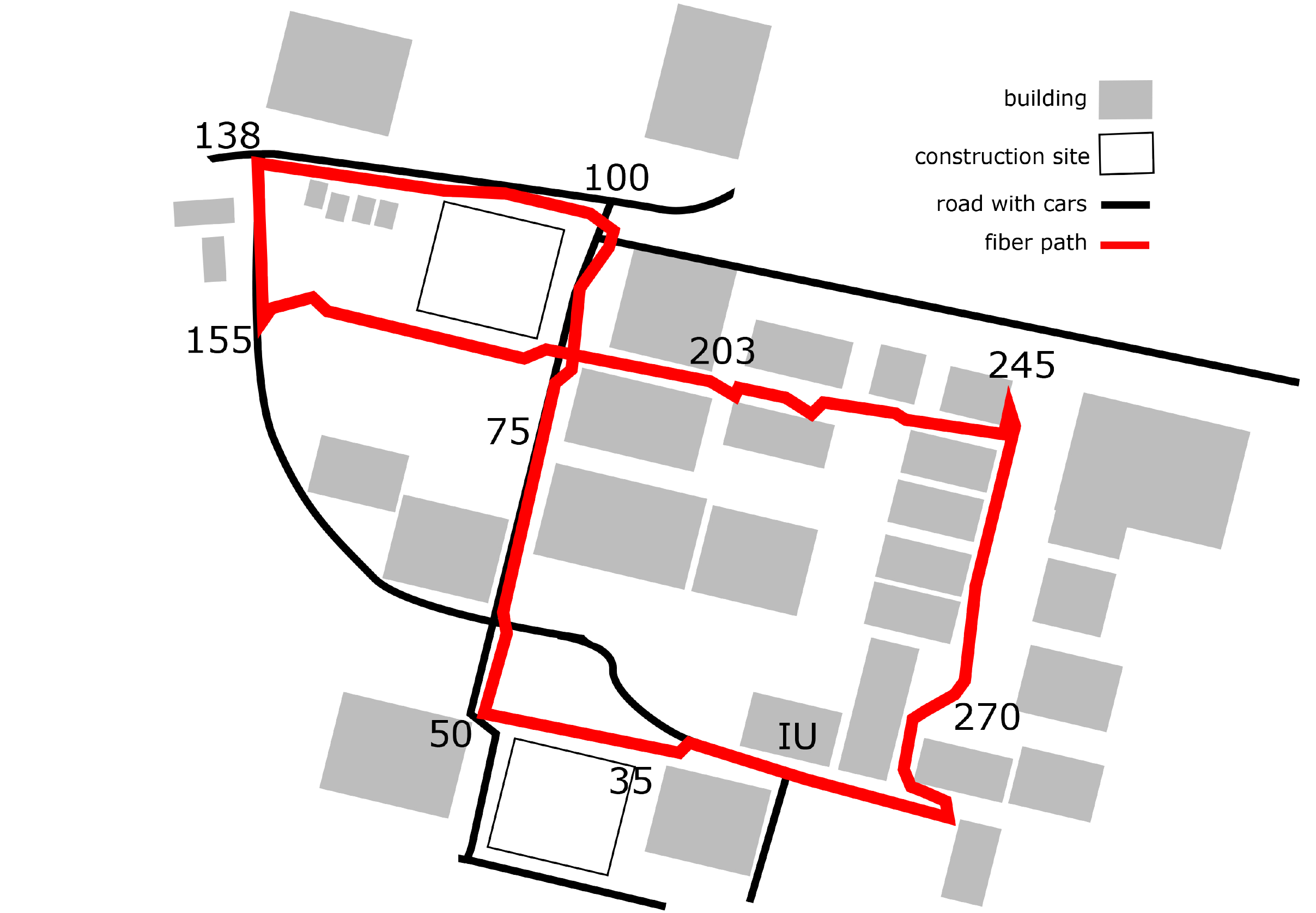}
        \hspace{0.5cm}
        \includegraphics[width=0.45\linewidth,trim={8.5cm 1cm 8.5cm 3cm},clip]{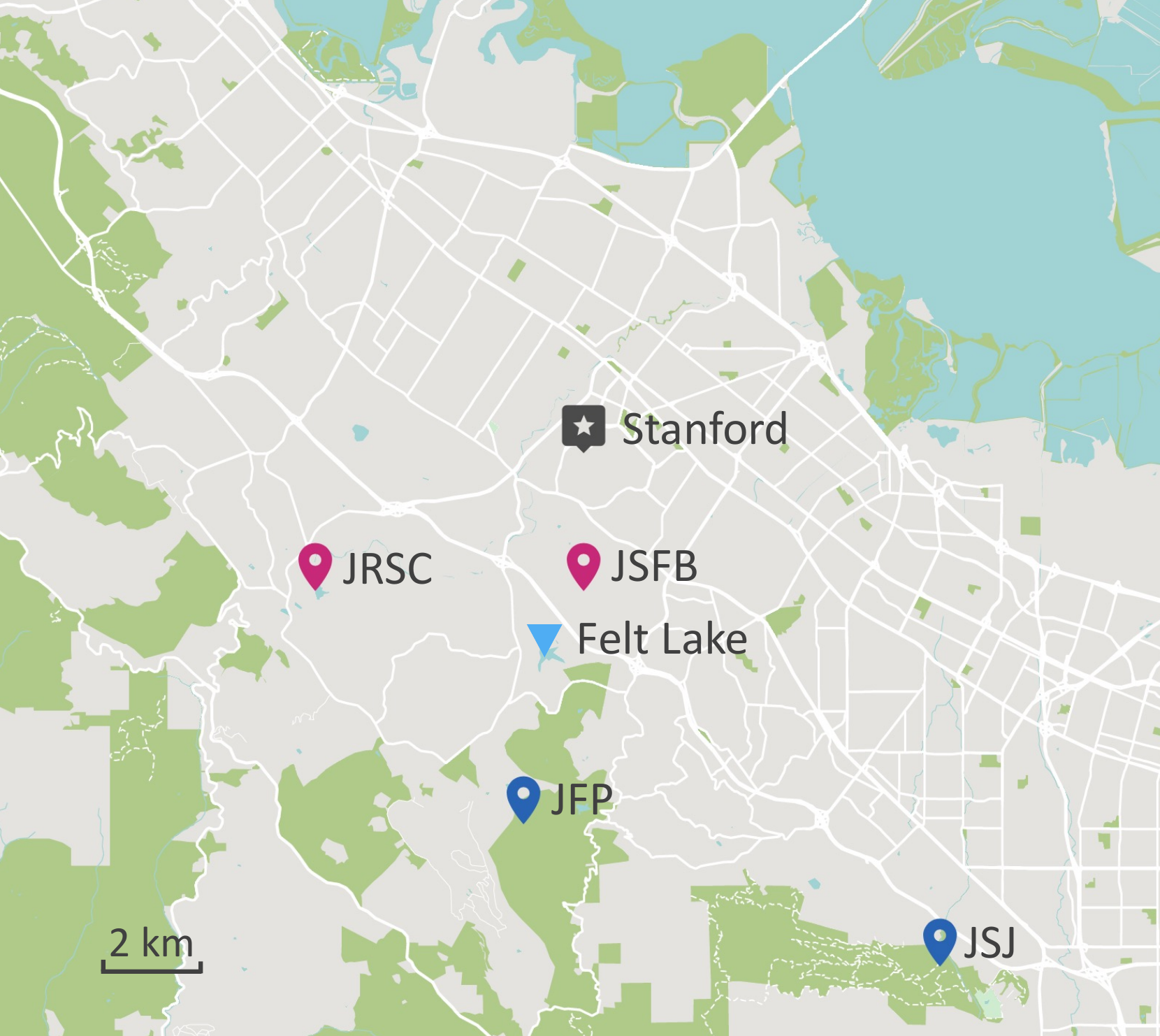}
    \end{center}
    \caption{(Left) Schematic map of the Stanford DAS array. ``IU'' indicates the location of the DAS interrogation unit. The array measures 600 m along its widest dimension. We number the recording channels along the cable. (Right) Map of continuously recording seismometers in the vicinity of the Stanford DAS array, marked by a black pin. The pink markers correspond to the two seismometer stations that we use in this study. 
    % Both are 3-component broadband seismometers. The Stanford Telescope, or JSFB station, is located 1.7 km away, while the Jasper Ridge Biological Preserve, or JRSC station, is located 5.8 km away. 
    The dark blue markers indicate the other seismometers that are within a 10 km range from Stanford.} %The blue inverted triangle marks the position of Felt Lake.}
    \label{fig:das_map}
\end{figure}

% The Stanford  DAS array is a figure-eight-shaped array of 2.4 km of fiber cables lying in the telecommunication conduits underneath the Stanford University campus (Figure~\ref{fig:das_map}). 
The Stanford DAS array is a figure-eight-shaped array of 2.4 km of fiber cables in the telecommunication conduits underneath the Stanford University campus (Figure~\ref{fig:das_map}). 
It measures 600 meters along its widest dimension.
We recorded the data continuously between 2016 and 2019 with a DAS interrogator unit (ODH3 from OptaSense) measuring strain rate at an 8 m channel spacing. The cables lie loosely in the conduits under a mix of natural and manufactured materials. Therefore, the quality of the signal and the coupling with the ground varies significantly throughout the array. The campus has many anthropogenic noise sources, with various roads with different traffic levels, quarry blasts within 15 km, plumbing and HVAC systems throughout, and multiple construction sites. Therefore, despite being in a suburban environment, this DAS array faces many of the same challenges as more urban acquisitions. The data is recorded at 50 samples per second, resulting in over 10 TB of data. Consequently, manual inspection of most data is infeasible, making automation tools critical to extracting the full value from the data. 

\begin{figure}
    \centering
        \includegraphics[width=0.8\linewidth,trim={0cm 0cm 0cm 1cm},clip]{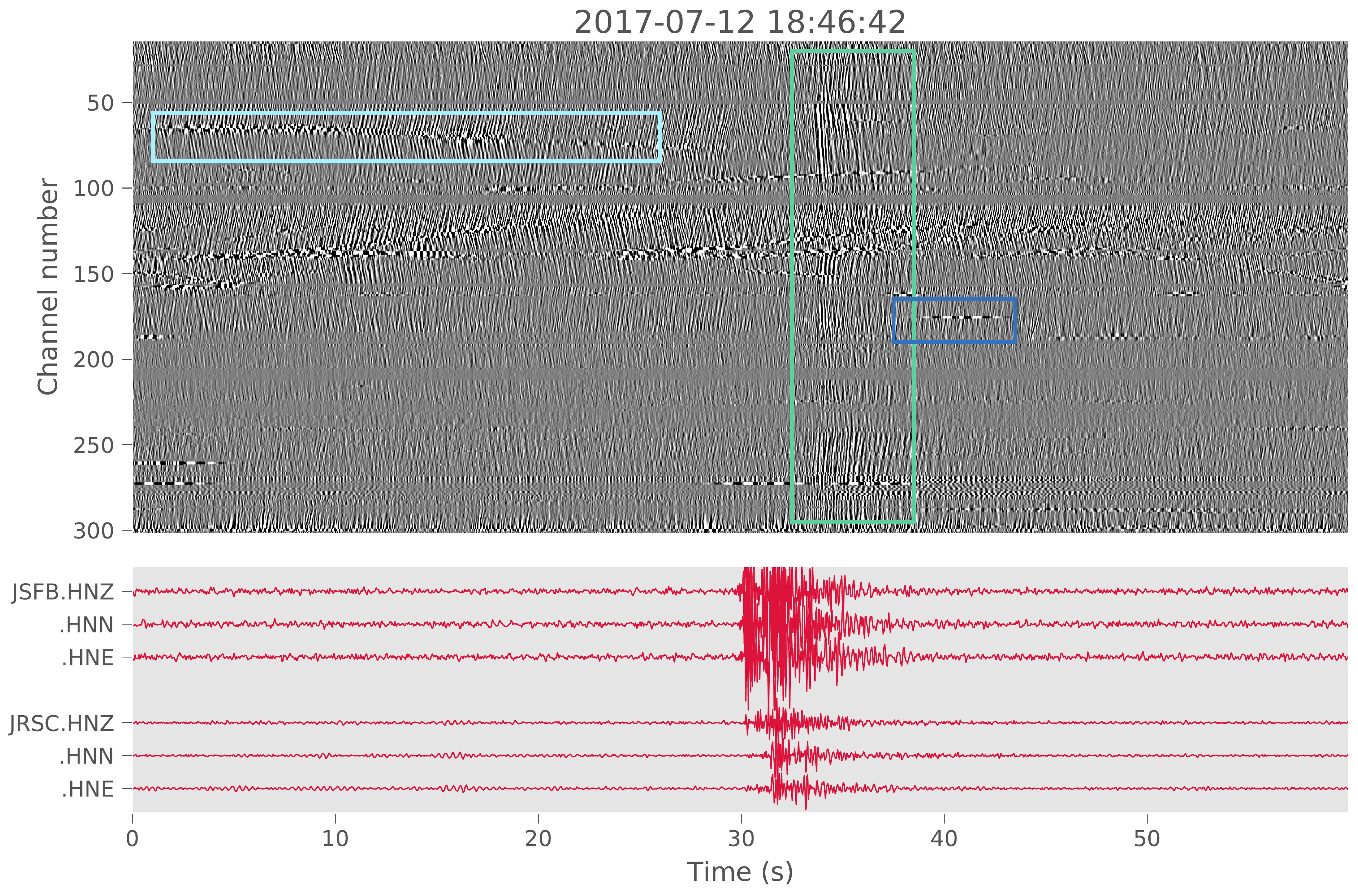}
    \caption{Example small-amplitude local earthquake recorded by the Stanford DAS array: an M 1.34 earthquake on July 12th, 2017, originating at Felt Lake, 5.3 km from the Stanford DAS array. The green box highlights the earthquake's arrival. The light blue box highlights an example of car noise; the dark blue box has some noise from a pumping or HVAC system. In red, we plot the data from the two nearest continuously recording seismometer stations, the JSFB and JRSC stations. HNZ denotes the vertical component, HNN and HNE denote the horizontal components.}
    \label{fig:raw_data1}
\end{figure}

% This fiber-optic DAS array detects a wide variety of seismic noise sources. It sits in a seismically active region, 20 km from the Pacific ocean, 7 km from the San Francisco bay, with highways on either side, a variety of roads with differing levels of traffic near the fiber, regular quarry blasts within 15 km, plumbing and HVAC systems throughout the site, multiple construction sites near the array, and foot and bicycle traffic throughout. 
% Figure~\ref{fig:raw_data1} presents an example of recorded data. 
% We band-pass filter the data between 1 and 12 Hz, a frequency band that contains most of the earthquake energy. 
The Stanford DAS array can capture the energy from local earthquakes. In Figure~\ref{fig:raw_data1}, we identify the arrival of the M 1.34 Felt Lake earthquake on July 12th, 2017 \cite{martin2017seismic}, originating at 5.3 km from the array. This observation confirms that we can repurpose existing infrastructure to monitor small-amplitude local earthquakes. We can validate the detection of local earthquakes with the data from the two nearest continuously recording seismometer stations (Figure~\ref{fig:das_map}). Both are 3-component broadband seismometers. The Stanford Telescope, or JSFB station, is located at 1.7 km of the Stanford array. The Jasper Ridge Biological Preserve, or JRSC station, is located at 5.8 km of the Stanford array. 
For small events close to the Stanford array, such as the Felt Lake earthquake presented in Figure \ref{fig:raw_data1}, only the nearby Stanford Telescope seismometer (located 4.8 km away from the Felt Lake earthquakes' epicenter) provides unambiguous validation of the event being an earthquake. The signal from the Jasper Ridge seismometer (located 7.1 km away from the Felt Lake earthquakes epicenter) is more ambiguous for this detection task.

% \begin{figure}
%     \begin{center}
%     %  trim={<left> <lower> <right> <upper>}
%         \includegraphics[width=0.7\linewidth,trim={4cm 0cm 4cm 0cm},clip]{fig/stations_map.pdf}
%     \end{center}
%     \caption{Map of continuously recording seismometers in the vicinity of the Stanford DAS array, marked by a black pin. The pink markers correspond to the two seismometer stations that we use in this study. 
%     % Both are 3-component broadband seismometers. The Stanford Telescope, or JSFB station, is located 1.7 km away, while the Jasper Ridge Biological Preserve, or JRSC station, is located 5.8 km away. 
%     The dark blue markers indicate the other seismometers that are within a 10 km range from Stanford. The blue inverted triangle marks the position of Felt Lake.}
%     \label{fig:stations_map}
% \end{figure}

In addition to the earthquake signals, we notice that the recorded DAS data contain many high-amplitude spurious noise types. The recorded amplitudes vary over different channels. Many noise patterns are associated with vehicle traffic, such as the one highlighted in light blue in Figure~\ref{fig:raw_data1}, typically of much stronger amplitude than the earthquake signal. 
% Since the data are filtered between 1 and 12 Hz, we cannot see the typical quasi-static signal generated by cars \cite{huot2017automatic,yuan2020near}, but we can distinguish the low frequency response as the vehicles move through the channels. 
We also observe frequent high amplitude noise isolated to single channels, such as the one highlighted in dark blue, probably due to pumps or HVAC systems. These limitations render the data unsuitable for conventional thresholding-based earthquake detection methods such as STA/LTA.

\section{Data processing}

To train an ML model for earthquake detection, we first create a curated dataset of earthquake and background noise examples. Using the Northern California Earthquake Data Center database \cite{ncedc2014northern}, we catalog all the events within 100 km over the period during which the Stanford DAS array was active and compute their expected arrival times. Our resulting catalog contains 3,925 earthquakes of various magnitudes and distances from the array, but we do not expect all these events to be detectable by the Stanford array. Therefore, we compute the expected local amplitudes of the events, as described by \citeA{lindsey2017fiber}, and only keep those above a certain threshold for the ML task. Based on visual inspection, we set 40 nano-strain as a reasonable threshold above which events are detectable from the DAS data. When compared to the events with cataloged arrivals at the nearest seismometer station, we note that this threshold captures the transition from detected to undetected events at the JSFB station. After removing events for which either the DAS data or one of the seismometers' data are missing, we obtain a catalog of 555 selected events. To provide noise examples for supervised learning, we randomly select background noise examples at random times at least 5 minutes away from any of the events in the catalog. We then visually inspect the noise catalog to ensure that none have apparent earthquake signals. 

We process the DAS data by computing the strain rate and removing the median over all channels for each time sample. We then band-pass filter both the DAS and the seismometer data between 1 and 12 Hz and decimate the data to 25 samples per second. The data have an extensive dynamic range, with strong amplitude spurious noise, so we clip the data to limit the impact of extreme values. We clip the DAS data at the 85th percentile and the seismometer data at the 98th percentile of absolute amplitudes. The clipping threshold of the DAS data is lower because the DAS data is overall noisier. We then divide the DAS data by its standard deviation over all channels and each seismometer component over its standard deviation. We process the recorded data shown in all the figures in this paper with the steps mentioned above to allow for visual comparisons between figures. 

We prepare 20.48 seconds (512 time-samples) data windows containing either earthquake signal or background noise. We extract them as larger 30-second windows and randomly crop them to the desired 20-second window size in the machine learning input pipeline. The random cropping ensures that the machine learning examples have events that occur with different time shifts. The resulting DAS data windows are 286 channels $\times$ 512 time-samples. The seismometer data windows are 512 $\times$ 6 components, with three components for each of the two seismometer stations. We split the data windows into a training, evaluation, and test set with respective ratios of 8:1:1.

\section{Machine learning models}

We frame the ML task as a binary classification between earthquake signal and background noise. We create two separate ML models to compare the potential for earthquake detection of fiber networks versus traditional sensors: a 2D CNN for detection on DAS data and a 1D CNN for seismometer recordings. 

We define each CNN as a modular sequence of convolution blocks, the characteristics of which are determined by hyperparameter tuning, as described by \cite[]{huot2022machine,huot2020detecting,huot2021detecting}. We define each convolution block as a sequence of a convolution layer, an activation function, and a downsampling layer. We add two fully-connected layers after the last convolution block. The hyperparameters related to network architecture are the number of convolution blocks, the number of filters, the type of activation \cite{ramachandran2017searching}, the type of downsampling, and the dropout rate \cite{srivastava2014dropout}. Hyperparameters related to training are batch size and learning rate. We optimize the network architecture and its training hyperparameters jointly by Bayesian optimization \cite{snoek2012practical}. We use the loss computed over the evaluation datasets as the performance metric. The networks are trained with Adam optimizer \cite{kingma2014adam} for 100 epochs (passes through the entire dataset) with early-stopping if there is no improvement after over 20 epochs. The training was performed on 4 NVIDIA V100 GPUs on an on-premise cluster.

Our best-performing 1D CNN and 2D CNN are remarkably similar, despite being optimized independently on separate datasets. The convolution layers have a 3 $\times$ 3 kernel and a 1 $\times$ 1 stride, while strided convolutions perform the downsampling with a 3 $\times$ 3 kernel and 2  $\times$ 2  stride (or their 1D equivalents for the 1D CNN). Both networks have 4 convolution blocks, use ReLU as the activation function, and have a 60\% dropout rate. The 1D CNN uses [8, 16, 32, 32] filters and the 2D CNN uses [32, 64, 128, 128] filters in its successive blocks. The 1D and 2D CNNs both use a learning rate of $10^{-3}$ and a batch size of 64 and 16, respectively.

\section{Experimental results}

We benchmark each model against its corresponding test dataset. The 2D CNN achieves 92.98\% accuracy on the DAS data (98.95\% area under the curve (AUC), 91.53\% precision, and 94.74\% recall). The 1D CNN achieves 95.54\% accuracy on the seismometer data (99.43\% AUC, 94.74\% precision, and 96.43\% recall). This result demonstrates that we can effectively detect earthquakes on both DAS and seismometer data, even when the earthquake amplitudes at arrival are similar to or lower than the background noise levels. Overall the 1D CNN achieves slightly better accuracy, which is expected since the seismometers have a better coupling with the ground and thus a better SNR. However, the model trained on DAS data performs quite well despite the fiber cables loosely lying in the telecommunication conduits and the strong amplitude urban noise. The 2D CNN makes full use of the additional dimension brought by the dense spatial sampling of DAS, allowing it to overcome the lower SNR in individual channels. The extensive hyperparameter search suggests that improving the detection accuracy beyond this point would require a larger labeled dataset.

\begin{figure}
    \begin{center}
        \includegraphics[width=0.49\linewidth]{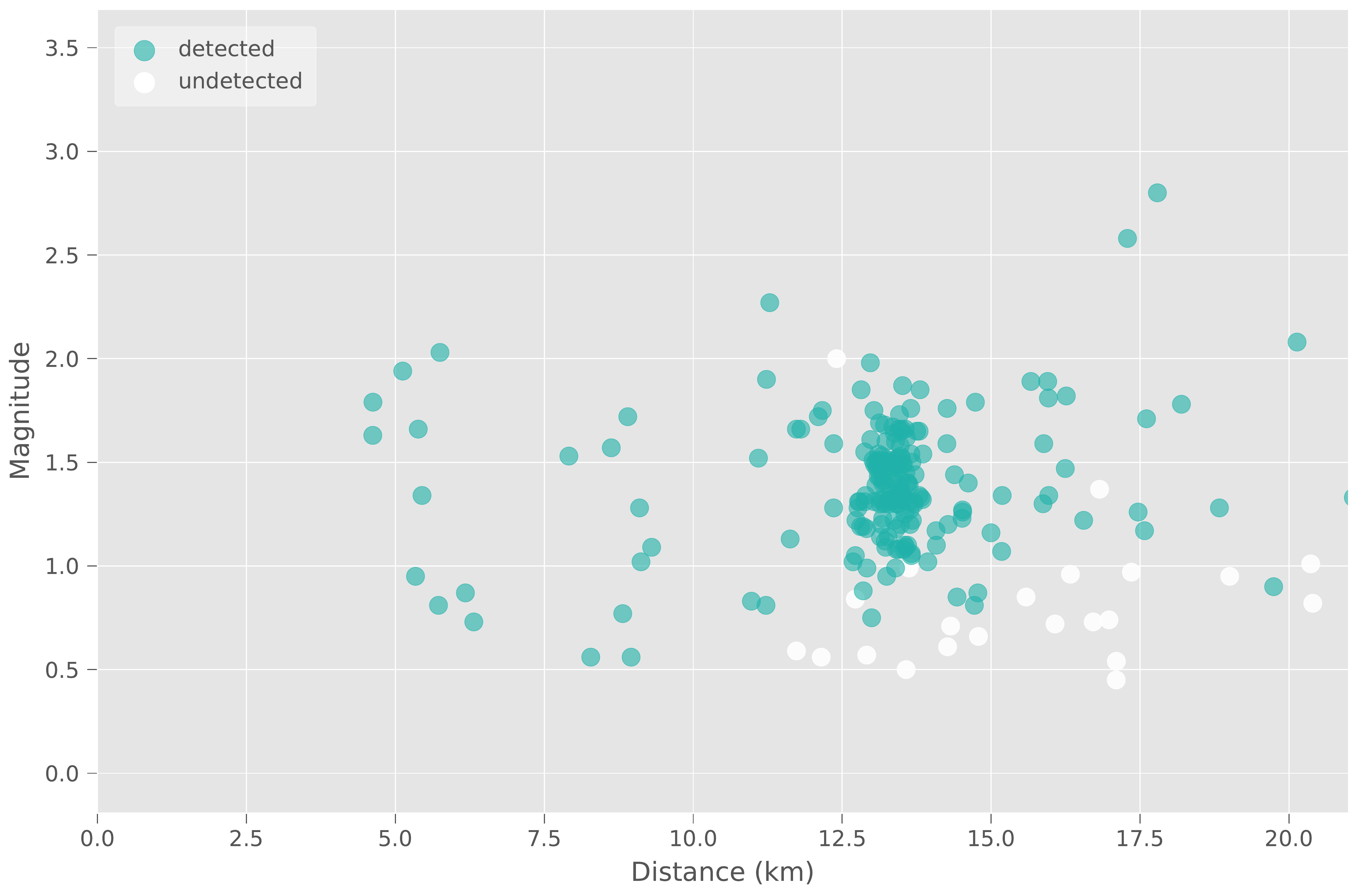}
        \includegraphics[width=0.49\linewidth]{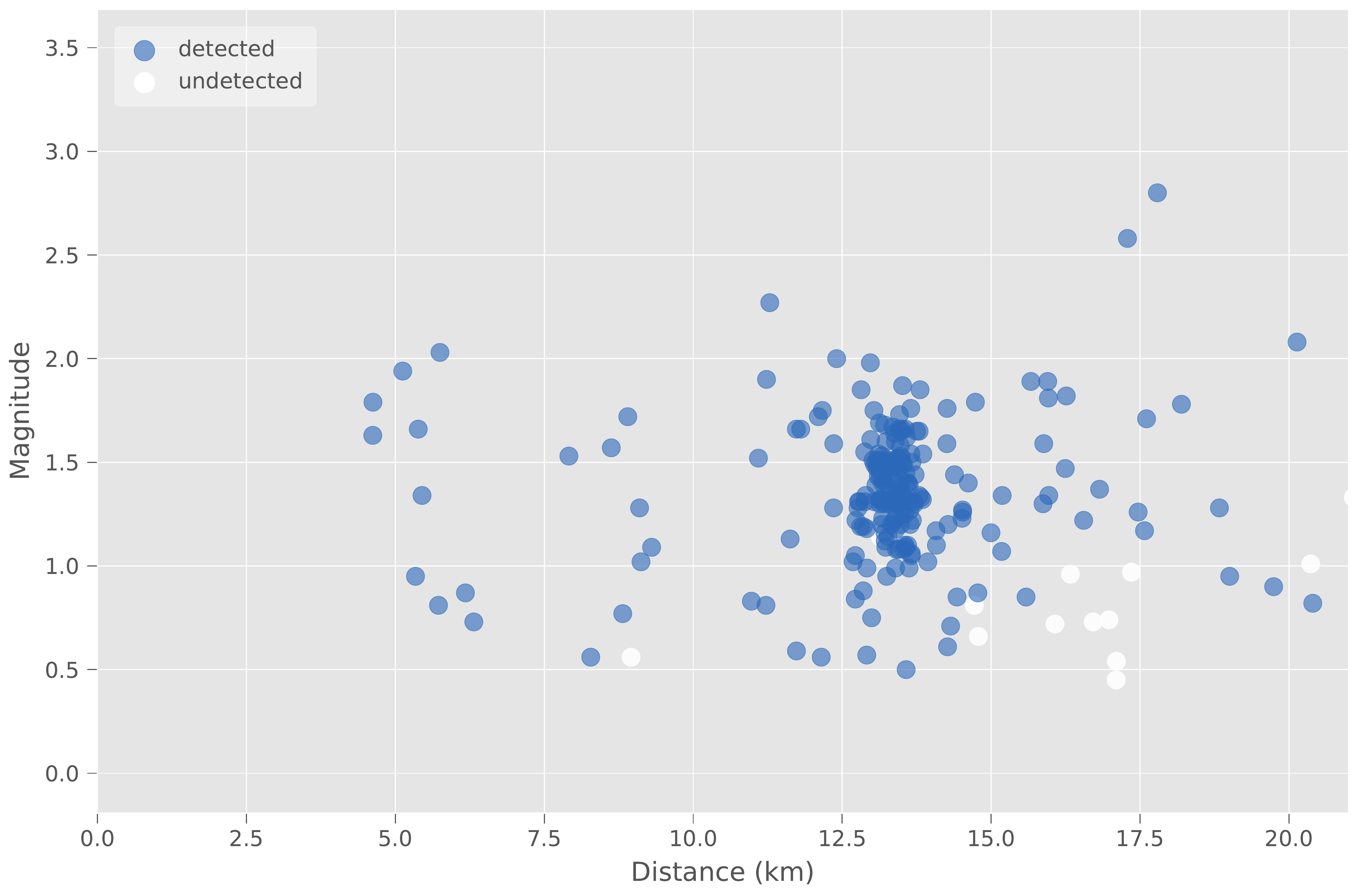}
    \end{center}
    \caption{Detected and undetected earthquakes within a 20 km radius by the Stanford array (Left) and the seismometer stations (Right).}
    \label{fig:all_events}
\end{figure}

When we test the trained models over all the events in our catalog, the detection results within a 20 km radius (Figure \ref{fig:all_events}) demonstrate that both acquisition methods perform well at detecting nearby earthquakes. In particular, the DAS array detected all events within close proximity ($<$ 10 km). The detection sensitivity for the DAS array and the seismometer stations decreases as anticipated with distance and lower magnitude.

We scale our detection workflow over three years of continuous data by sliding a 20-second window through the recordings with a 10-second stride and classifying each frame independently. One day of continuous data corresponds to about 8,000 data windows, while a year of recordings represents over 3 million. These numbers illustrate the difficulty of the detection task on continuous data, as minor classification errors can quickly translate to thousands of misclassifications when scaled over time. The 2D CNN on the DAS data classifies 169,510 windows as events over three years (2.2\% of windows), while the 1D CNN on the seismometer data flags about 33,604 windows (0.4\% of windows). While these absolute numbers may seem high, they only represent a small percentage of the total number of data windows, which is in line with the high accuracy of the CNNs on the benchmark data. However, this initial result suggests that we still have many false detections.

From visual inspection, many of the false detections on the DAS and the seismometer data are triggered by noise modes specific to each sensor. The false detections on the DAS data are due to noise that reverberates across all channels, such as construction work on campus or large vehicles hitting a bump on the road. In contrast, the ones on the seismometers are triggered by spurious impulsive noise. These types of noise do not come from the same sources and, more importantly, do not occur simultaneously. Therefore, we take the intersection of the detections from both types of sensors to validate events, significantly reducing the false detection rate. Using this approach, we narrow the detections to 2,635 windows, which illustrates how fiber acquisitions can complement sparse seismometer networks. In particular, despite having lesser coupling with the ground and lower SNR, they record a different type of data, making them susceptible to very different types of noise than seismometers. This specificity makes them valuable for cross-validating events across other sensors.

We consolidate the 2,635 detections by combining overlapping windows corresponding to the same event, leaving us with 1,879 detected events. The majority of these 1,879 detections correspond to events cataloged in the NDEDC database. To investigate whether these events could be uncataloged local small-amplitude events, we remove all the events that coincide with arrivals in the NDEDC catalog. 

\begin{figure}
    \centering
        \includegraphics[width=0.8\linewidth,trim={0cm 0cm 0cm 1cm},clip]{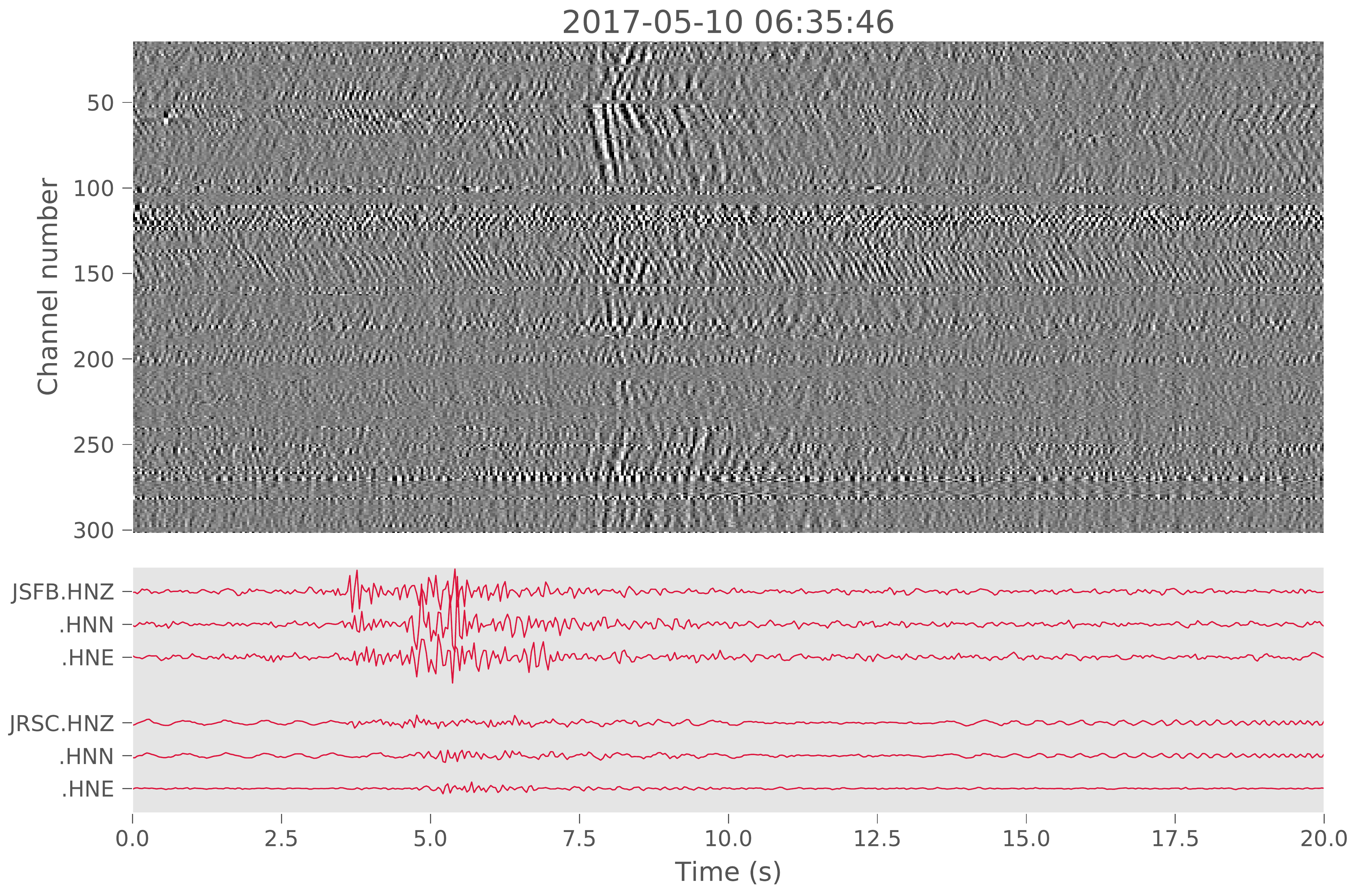}
\caption{Instance on 5/10/2017 of the local repeating Felt Lake earthquake that does not figure in the NCEDC database but was detected by our detection workflow.}
\label{fig:uncataloged_feltlake1}
\end{figure}

\begin{figure}
    \centering
    \includegraphics[width=0.45\linewidth]{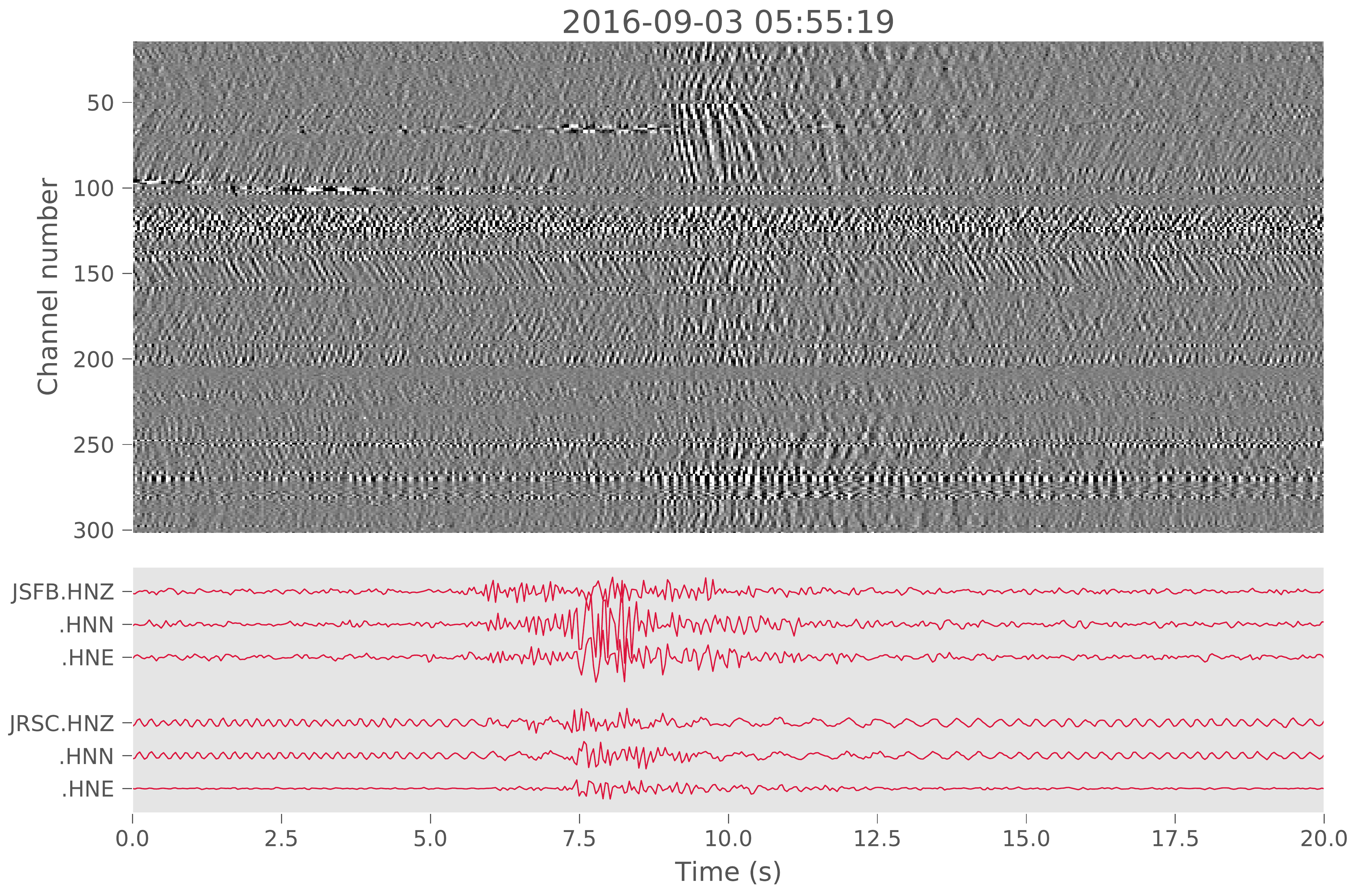}
    \includegraphics[width=0.45\linewidth]{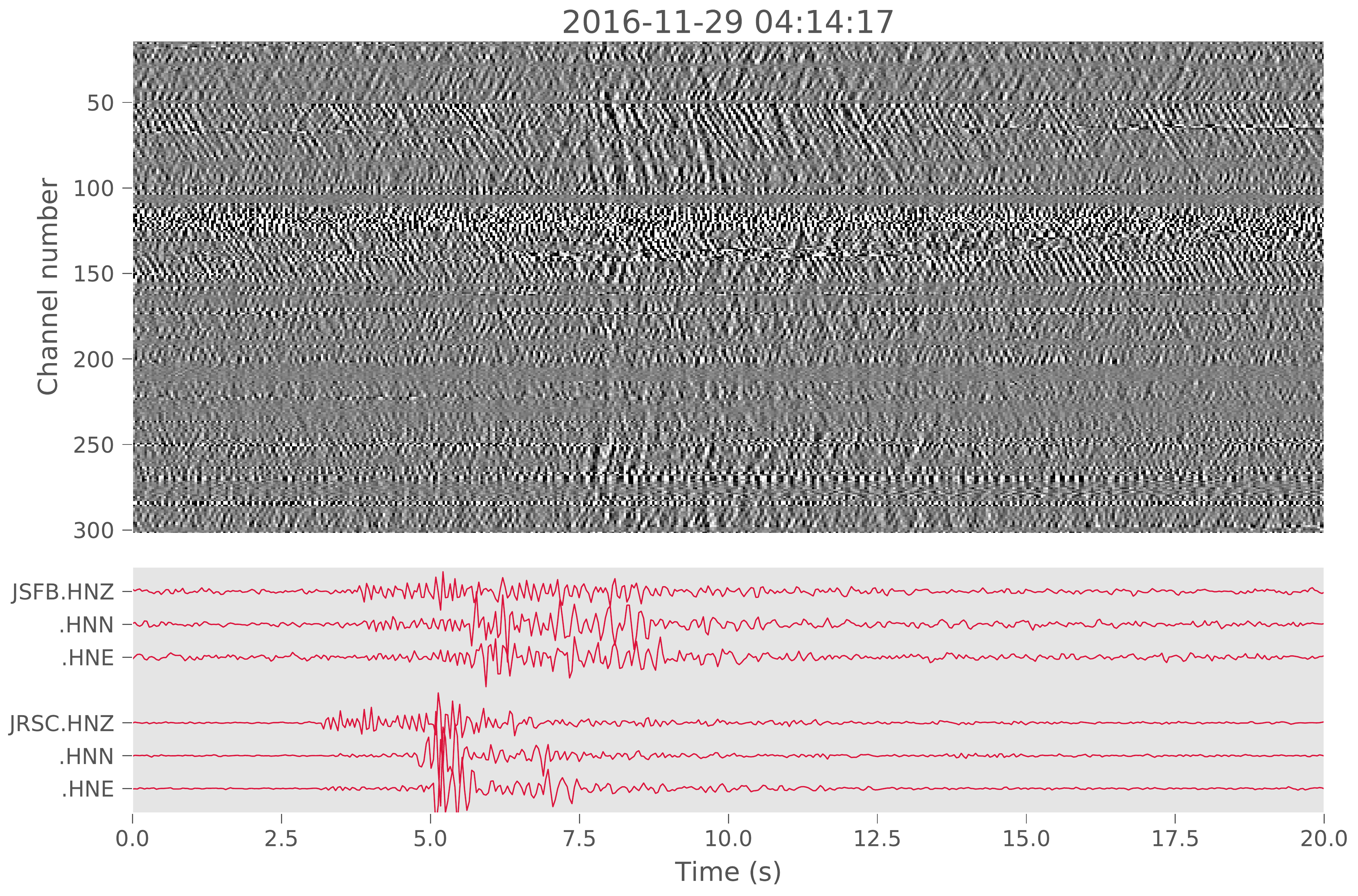}
    \includegraphics[width=0.45\linewidth]{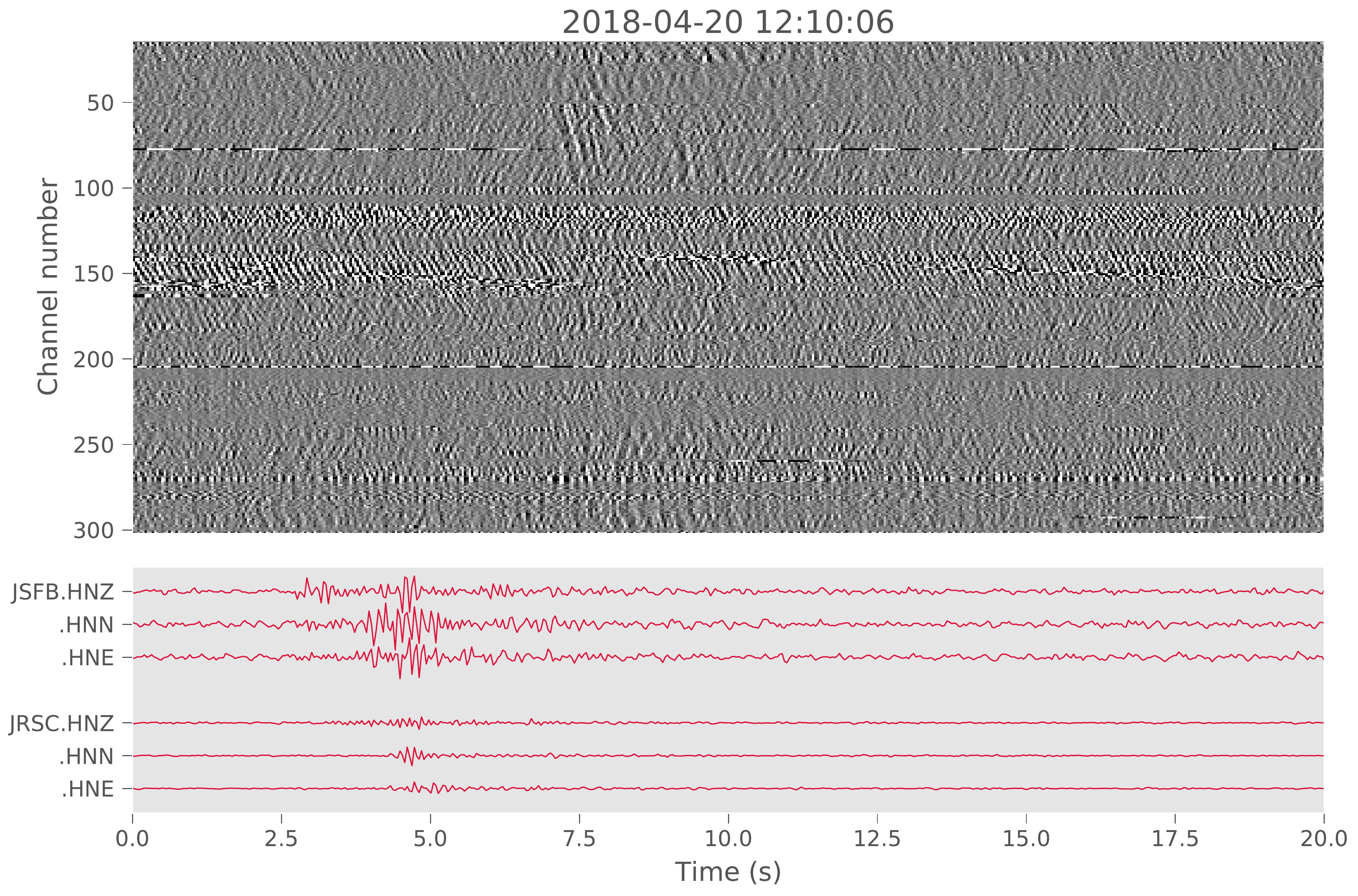}
    \includegraphics[width=0.45\linewidth]{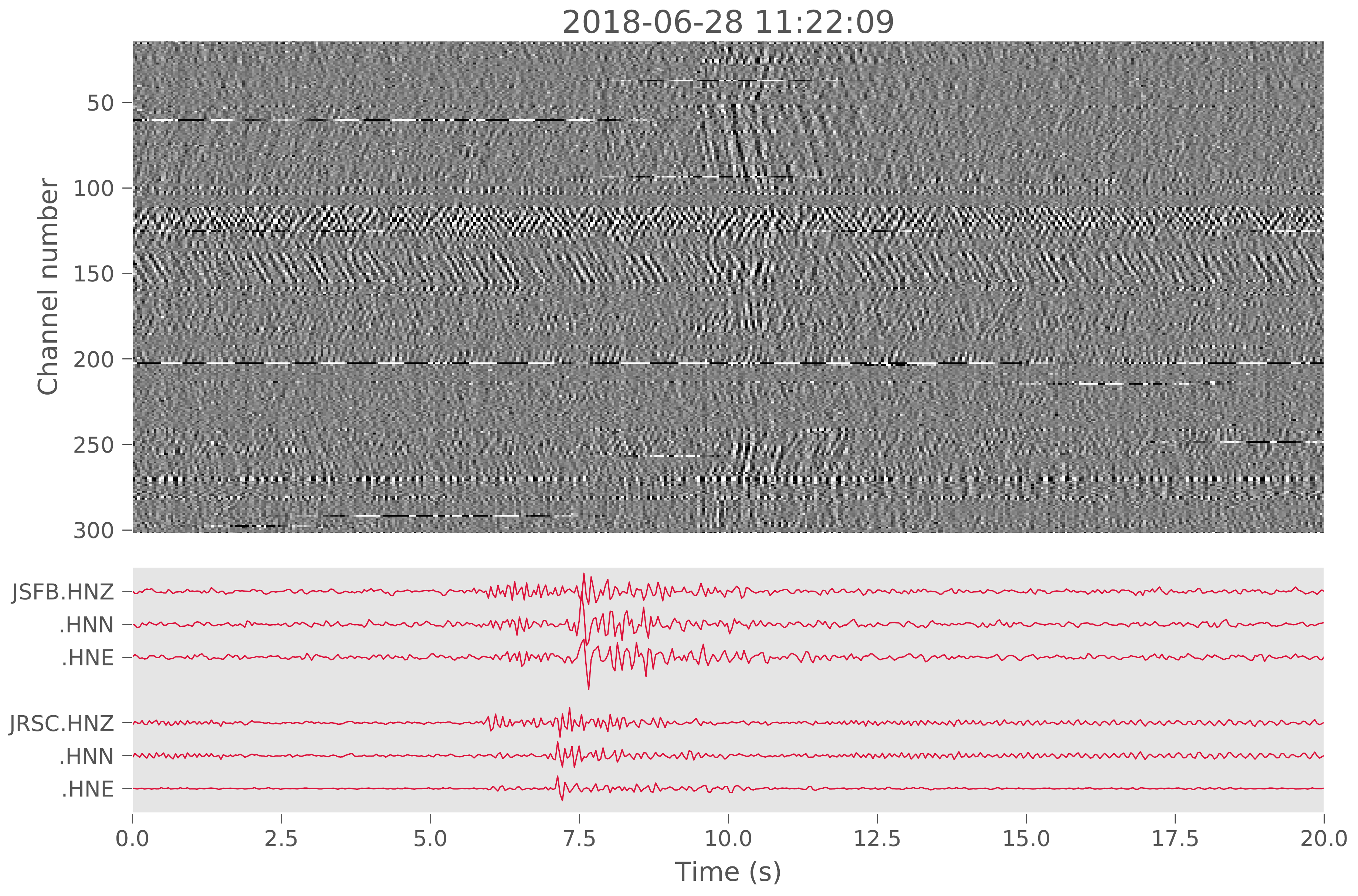}
\caption{Examples of uncataloged earthquakes detected by our detection workflow. The time at the top of each subfigure is indicated in UTC.}
\label{fig:uncataloged_events}
\end{figure}

After removing the previously-cataloged events, we have 181 detections over the three years of continuous data. After visual inspection of these candidate events, some correspond to the coda of earthquakes further away, while others look like noise. Our CNN did not separate nearby earthquakes perfectly but narrowed down the numbers sufficiently to validate the remaining events manually. We identify about ten detections as potential uncataloged local small-amplitude earthquakes. In particular, our detection workflow successfully retrieves two instances of the local repeating Felt Lake earthquake that does not figure in the NCEDC database but was identified by cross-correlation in a previous study \cite{biondi2021using} (Figure \ref{fig:uncataloged_feltlake1}). Detecting small earthquakes by cross-correlation requires a previously cataloged event originating at the same location to create pattern-matching templates. Our machine-learning detection methodology does not require a previously-established template nor does it require events to repeat. The task of finding these small previously-uncataloged events from 3 years of continuous recordings would not have been possible without automated processing methods. Furthermore, the detected events could be used as templates for template matching, and likely identify even more small uncatalogued events. Figure \ref{fig:uncataloged_events} shows examples of likely candidates for this template matching procedure.

% \begin{figure}
%     \centering
%         \includegraphics[width=\linewidth]{fig/window_20170713_055610_crop.pdf}
% \caption{Example of uncataloged earthquake on 7/1fig/2017 detected by our detection workflow: Zoomed-in 20-second data window with the detected event. The time at the top of the figure is indicated in UTC. All recorded data are plotted with the same processing steps and on the same scale throughout this study to allow visual comparisons.}
% \label{fig:uncataloged_feltlake2}
% \end{figure}

\section{Conclusion}
We demonstrate that we can detect local earthquakes using the fiber cables laying in the telecommunication conduits under the Stanford University campus. We compare the detection capabilities of our fiber array versus neighboring seismometers. To train the ML model, we create a curated dataset of earthquake events and background noise for both types of sensors. We train CNNs to perform earthquake detection on each data type and show that the models successfully identify local earthquakes, even when the earthquake signal is weak, despite the high-amplitude anthropogenic noise in the fiber data. Both sensors perform well at detecting local earthquakes within 10 km.

Scaling the detection workflow over three years of continuous data reveals that combining both types of acquisition for validating events significantly lowers the false detections rate. Our automated processing workflow helps us identify a small number of candidate local events among the detections. After excluding events that are already present in the NDEDC database, manual analysis of the remaining events reveals some uncataloged small-amplitude local earthquakes. A small local repeating earthquake identified by our workflow corresponds to events identified by cross-correlation in a previous study \cite{biondi2021using}. Unlike cross-correlation, our machine-learning detection does not require a previously-cataloged event to use as a template nor requires events to repeat. Additionally, once the ML model is trained, it can perform detections on continuous data in a matter of seconds, allowing us to scale the detections to large data volumes. The task of finding these small uncataloged events from 3 years of continuous recordings would not have been possible without automated processing methods. This result suggests that repurposing infrastructure fiber cables combined with ML processing can complement sparse seismometer networks in urban areas. This type of data acquisition would be particularly valuable in areas with little seismometer coverage.

\acknowledgments

We thank OptaSense for making available their ODH3 interrogator unit. We would like to thank Stanford IT for their help installing the array and Stanford Center for Computational Earth and Environmental Sciences for their computing resources. We thank the Stanford Exploration Project affiliate members for their financial support. 

\section*{Open research}

The seismometer data used in this study are available in the Northern California Earthquake Data Center database \cite{ncedc2014northern}, using network codes BK and NC. The Stanford DAS data are available at:

\noindent https://premonition.stanford.edu/fantine/stanford-das-array \cite{fantinehuot20226344570}.

\noindent We implement the CNN using TensorFlow \cite[]{tensorflow2015-whitepaper} and use the GPyOpt library \cite[]{gpyopt2016} for the Bayesian optimization. The source code associated with this manuscript is available in the following repository:

\noindent https://github.com/fantine/earthquake-detection-ml \cite{fantinehuot20226344771}.

% This section is optional. Include any Acknowledgments here.
% The acknowledgments should list:\\
% All funding sources related to this work from all authors\\
% Any real or perceived financial conflicts of interests for any author\\
% Other affiliations for any author that may be perceived as having a conflict of interest with respect to the results of this paper.\\
% It is also the appropriate place to thank colleagues and other contributors. AGU does not normally allow dedications.

%% ------------------------------------------------------------------------ %%
%% References and Citations

%%%%%%%%%%%%%%%%%%%%%%%%%%%%%%%%%%%%%%%%%%%%%%%
%
\bibliography{references} %don't specify the file extension

\begin{thebibliography}{}

\bibitem [\protect \citeauthoryear {%
Abadi%
\ \protect \BOthers {.}}{%
Abadi%
\ \protect \BOthers {.}}{%
{\protect \APACyear {2015}}%
}]{%
tensorflow2015-whitepaper}
\APACinsertmetastar {%
tensorflow2015-whitepaper}%
\begin{APACrefauthors}%
Abadi, M.%
, Agarwal, A.%
, Barham, P.%
, Brevdo, E.%
, Chen, Z.%
, Citro, C.%
\BDBL {}Zheng, X.%
\end{APACrefauthors}%
\unskip\
\newblock
\APACrefYearMonthDay{2015}{}{}.
\newblock
\APACrefbtitle {{TensorFlow}: Large-scale machine learning on heterogeneous
  systems.} {{TensorFlow}: Large-scale machine learning on heterogeneous
  systems.}
\newblock
\begin{APACrefURL} \url{https://www.tensorflow.org/} \end{APACrefURL}
\newblock
\APACrefnote{Software available from tensorflow.org}
\PrintBackRefs{\CurrentBib}

\bibitem [\protect \citeauthoryear {%
Bakku%
}{%
Bakku%
}{%
{\protect \APACyear {2015}}%
}]{%
bakku2015fracture}
\APACinsertmetastar {%
bakku2015fracture}%
\begin{APACrefauthors}%
Bakku, S\BPBI K.%
\end{APACrefauthors}%
\unskip\
\newblock
\APACrefYear{2015}.
\unskip\
\newblock
\APACrefbtitle {Fracture characterization from seismic measurements in a
  borehole} {Fracture characterization from seismic measurements in a
  borehole}\ \APACtypeAddressSchool {\BUPhD}{}{}.
\unskip\
\newblock
\APACaddressSchool {}{Massachusetts Institute of Technology}.
\PrintBackRefs{\CurrentBib}

\bibitem [\protect \citeauthoryear {%
Biondi%
, Clapp%
, Yuan%
\BCBL {}\ \BBA {} Huot%
}{%
Biondi%
, Clapp%
\BCBL {}\ \protect \BOthers {.}}{%
{\protect \APACyear {2021}}%
}]{%
biondi2021scaling}
\APACinsertmetastar {%
biondi2021scaling}%
\begin{APACrefauthors}%
Biondi, B\BPBI L.%
, Clapp, R\BPBI G.%
, Yuan, S.%
\BCBL {}\ \BBA {} Huot, F.%
\end{APACrefauthors}%
\unskip\
\newblock
\APACrefYearMonthDay{2021}{}{}.
\newblock
{\BBOQ}\APACrefatitle {Scaling up to city-wide dark-fiber seismic arrays:
  Lessons from five years of the Stanford DAS array project} {Scaling up to
  city-wide dark-fiber seismic arrays: Lessons from five years of the stanford
  das array project}.{\BBCQ}
\newblock
\BIn{} \APACrefbtitle {First International Meeting for Applied Geoscience \&
  Energy} {First international meeting for applied geoscience \& energy}\
  (\BPGS\ 3225--3229).
\PrintBackRefs{\CurrentBib}

\bibitem [\protect \citeauthoryear {%
Biondi%
, Yuan%
, Martin%
, Huot%
\BCBL {}\ \BBA {} Clapp%
}{%
Biondi%
, Yuan%
\BCBL {}\ \protect \BOthers {.}}{%
{\protect \APACyear {2021}}%
}]{%
biondi2021using}
\APACinsertmetastar {%
biondi2021using}%
\begin{APACrefauthors}%
Biondi, B\BPBI L.%
, Yuan, S.%
, Martin, E\BPBI R.%
, Huot, F.%
\BCBL {}\ \BBA {} Clapp, R\BPBI G.%
\end{APACrefauthors}%
\unskip\
\newblock
\APACrefYearMonthDay{2021}{}{}.
\newblock
{\BBOQ}\APACrefatitle {Using telecommunication fiber infrastructure for
  earthquake monitoring and near-surface characterization} {Using
  telecommunication fiber infrastructure for earthquake monitoring and
  near-surface characterization}.{\BBCQ}
\newblock
\APACjournalVolNumPages{Distributed Acoustic Sensing in Geophysics: Methods and
  Applications}{}{}{131--148}.
\PrintBackRefs{\CurrentBib}

\bibitem [\protect \citeauthoryear {%
Daley%
\ \protect \BOthers {.}}{%
Daley%
\ \protect \BOthers {.}}{%
{\protect \APACyear {2013}}%
}]{%
daley2013field}
\APACinsertmetastar {%
daley2013field}%
\begin{APACrefauthors}%
Daley, T\BPBI M.%
, Freifeld, B\BPBI M.%
, Ajo-Franklin, J.%
, Dou, S.%
, Pevzner, R.%
, Shulakova, V.%
\BDBL {}others%
\end{APACrefauthors}%
\unskip\
\newblock
\APACrefYearMonthDay{2013}{}{}.
\newblock
{\BBOQ}\APACrefatitle {Field testing of fiber-optic distributed acoustic
  sensing (DAS) for subsurface seismic monitoring} {Field testing of
  fiber-optic distributed acoustic sensing (das) for subsurface seismic
  monitoring}.{\BBCQ}
\newblock
\APACjournalVolNumPages{The Leading Edge}{32}{6}{699--706}.
\PrintBackRefs{\CurrentBib}

\bibitem [\protect \citeauthoryear {%
Field%
\ \protect \BOthers {.}}{%
Field%
\ \protect \BOthers {.}}{%
{\protect \APACyear {2015}}%
}]{%
field2015ucerf3}
\APACinsertmetastar {%
field2015ucerf3}%
\begin{APACrefauthors}%
Field, E\BPBI H.%
\BCBT {}\ \BOthersPeriod {.}
\end{APACrefauthors}%
\unskip\
\newblock
\APACrefYearMonthDay{2015}{}{}.
\newblock
\APACrefbtitle {UCERF3: A new earthquake forecast for California's complex
  fault system} {Ucerf3: A new earthquake forecast for california's complex
  fault system}\ \APACbVolEdTR{}{\BTR{}}.
\newblock
\APACaddressInstitution{}{US Geological Survey}.
\PrintBackRefs{\CurrentBib}

\bibitem [\protect \citeauthoryear {%
Huot%
}{%
Huot%
}{%
{\protect \APACyear {2022}}%
{\protect \APACexlab {{\protect \BCnt {1}}}}}]{%
fantinehuot20226344771}
\APACinsertmetastar {%
fantinehuot20226344771}%
\begin{APACrefauthors}%
Huot, F.%
\end{APACrefauthors}%
\unskip\
\newblock
\APACrefYearMonthDay{2022{\protect \BCnt {1}}}{}{}.
\newblock
\APACrefbtitle {{Earthquake detection via fiber-optic cables using deep
  learning}.} {{Earthquake detection via fiber-optic cables using deep
  learning}.}
\newblock
\APACaddressPublisher{}{Zenodo}.
\newblock
\begin{APACrefURL} \url{https://doi.org/10.5281/zenodo.6344771}
  \end{APACrefURL}
\newblock
\begin{APACrefDOI} \doi{10.5281/zenodo.6344771} \end{APACrefDOI}
\PrintBackRefs{\CurrentBib}

\bibitem [\protect \citeauthoryear {%
Huot%
}{%
Huot%
}{%
{\protect \APACyear {2022}}%
{\protect \APACexlab {{\protect \BCnt {2}}}}}]{%
huot2022machine}
\APACinsertmetastar {%
huot2022machine}%
\begin{APACrefauthors}%
Huot, F.%
\end{APACrefauthors}%
\unskip\
\newblock
\APACrefYear{2022{\protect \BCnt {2}}}.
\unskip\
\newblock
\APACrefbtitle {Machine Learning for Seismic Event Detection. A Story in Three
  Parts: Earthquakes, Microseismic, and Tectonic Tremors} {Machine learning for
  seismic event detection. a story in three parts: Earthquakes, microseismic,
  and tectonic tremors}\ \APACtypeAddressSchool {\BUPhD}{}{}.
\unskip\
\newblock
\APACaddressSchool {}{Stanford University}.
\PrintBackRefs{\CurrentBib}

\bibitem [\protect \citeauthoryear {%
Huot%
}{%
Huot%
}{%
{\protect \APACyear {2022}}%
{\protect \APACexlab {{\protect \BCnt {3}}}}}]{%
fantinehuot20226344570}
\APACinsertmetastar {%
fantinehuot20226344570}%
\begin{APACrefauthors}%
Huot, F.%
\end{APACrefauthors}%
\unskip\
\newblock
\APACrefYearMonthDay{2022{\protect \BCnt {3}}}{}{}.
\newblock
\APACrefbtitle {{Stanford fiber-optic DAS array: Earthquake detection
  dataset}.} {{Stanford fiber-optic DAS array: Earthquake detection dataset}.}
\newblock
\APACaddressPublisher{}{Zenodo}.
\newblock
\begin{APACrefURL} \url{https://doi.org/10.5281/zenodo.6344570}
  \end{APACrefURL}
\newblock
\begin{APACrefDOI} \doi{10.5281/zenodo.6344570} \end{APACrefDOI}
\PrintBackRefs{\CurrentBib}

\bibitem [\protect \citeauthoryear {%
Huot%
\ \BBA {} Biondi%
}{%
Huot%
\ \BBA {} Biondi%
}{%
{\protect \APACyear {2018}}%
}]{%
huot2018machine}
\APACinsertmetastar {%
huot2018machine}%
\begin{APACrefauthors}%
Huot, F.%
\BCBT {}\ \BBA {} Biondi, B\BPBI L.%
\end{APACrefauthors}%
\unskip\
\newblock
\APACrefYearMonthDay{2018}{}{}.
\newblock
{\BBOQ}\APACrefatitle {Machine learning algorithms for automated seismic
  ambient noise processing applied to DAS acquisition} {Machine learning
  algorithms for automated seismic ambient noise processing applied to das
  acquisition}.{\BBCQ}
\newblock
\BIn{} \APACrefbtitle {SEG Technical Program Expanded Abstracts 2018} {Seg
  technical program expanded abstracts 2018}\ (\BPGS\ 5501--5505).
\newblock
\APACaddressPublisher{}{Society of Exploration Geophysicists}.
\PrintBackRefs{\CurrentBib}

\bibitem [\protect \citeauthoryear {%
Huot%
\ \BBA {} Biondi%
}{%
Huot%
\ \BBA {} Biondi%
}{%
{\protect \APACyear {2020}}%
}]{%
huot2020detecting}
\APACinsertmetastar {%
huot2020detecting}%
\begin{APACrefauthors}%
Huot, F.%
\BCBT {}\ \BBA {} Biondi, B\BPBI L.%
\end{APACrefauthors}%
\unskip\
\newblock
\APACrefYearMonthDay{2020}{}{}.
\newblock
{\BBOQ}\APACrefatitle {Detecting earthquakes through telecom fiber using a
  convolutional neural network} {Detecting earthquakes through telecom fiber
  using a convolutional neural network}.{\BBCQ}
\newblock
\BIn{} \APACrefbtitle {SEG Technical Program Expanded Abstracts 2020} {Seg
  technical program expanded abstracts 2020}\ (\BPGS\ 3452--3456).
\newblock
\APACaddressPublisher{}{Society of Exploration Geophysicists}.
\PrintBackRefs{\CurrentBib}

\bibitem [\protect \citeauthoryear {%
Huot%
, Biondi%
\BCBL {}\ \BBA {} Beroza%
}{%
Huot%
, Biondi%
\BCBL {}\ \BBA {} Beroza%
}{%
{\protect \APACyear {2018}}%
}]{%
huot2018jump}
\APACinsertmetastar {%
huot2018jump}%
\begin{APACrefauthors}%
Huot, F.%
, Biondi, B\BPBI L.%
\BCBL {}\ \BBA {} Beroza, G.%
\end{APACrefauthors}%
\unskip\
\newblock
\APACrefYearMonthDay{2018}{}{}.
\newblock
{\BBOQ}\APACrefatitle {Jump-starting neural network training for seismic
  problems} {Jump-starting neural network training for seismic
  problems}.{\BBCQ}
\newblock
\BIn{} \APACrefbtitle {2018 SEG International Exposition and Annual Meeting.}
  {2018 seg international exposition and annual meeting.}
\PrintBackRefs{\CurrentBib}

\bibitem [\protect \citeauthoryear {%
Huot%
\ \protect \BOthers {.}}{%
Huot%
\ \protect \BOthers {.}}{%
{\protect \APACyear {2021}}%
}]{%
huot2021detecting}
\APACinsertmetastar {%
huot2021detecting}%
\begin{APACrefauthors}%
Huot, F.%
, Lellouch, A.%
, Given, P.%
, Clapp, R\BPBI G.%
, Biondi, B\BPBI L.%
, Nemeth, T.%
\BCBL {}\ \BBA {} Nihei, K.%
\end{APACrefauthors}%
\unskip\
\newblock
\APACrefYearMonthDay{2021}{}{}.
\newblock
{\BBOQ}\APACrefatitle {Detecting microseismic events on DAS fiber with
  super-human accuracy} {Detecting microseismic events on das fiber with
  super-human accuracy}.{\BBCQ}
\newblock
\BIn{} \APACrefbtitle {SEG/AAPG/SEPM First International Meeting for Applied
  Geoscience \& Energy.} {Seg/aapg/sepm first international meeting for applied
  geoscience \& energy.}
\PrintBackRefs{\CurrentBib}

\bibitem [\protect \citeauthoryear {%
Huot%
, Ma%
, Cieplicki%
, Martin%
\BCBL {}\ \BBA {} Biondi%
}{%
Huot%
\ \protect \BOthers {.}}{%
{\protect \APACyear {2017}}%
}]{%
huot2017automatic}
\APACinsertmetastar {%
huot2017automatic}%
\begin{APACrefauthors}%
Huot, F.%
, Ma, Y.%
, Cieplicki, R.%
, Martin, E.%
\BCBL {}\ \BBA {} Biondi, B.%
\end{APACrefauthors}%
\unskip\
\newblock
\APACrefYearMonthDay{2017}{}{}.
\newblock
{\BBOQ}\APACrefatitle {Automatic noise exploration in urban areas} {Automatic
  noise exploration in urban areas}.{\BBCQ}
\newblock
\BIn{} \APACrefbtitle {SEG Technical Program Expanded Abstracts 2017} {Seg
  technical program expanded abstracts 2017}\ (\BPGS\ 5027--5032).
\newblock
\APACaddressPublisher{}{Society of Exploration Geophysicists}.
\PrintBackRefs{\CurrentBib}

\bibitem [\protect \citeauthoryear {%
Huot%
, Martin%
\BCBL {}\ \BBA {} Biondi%
}{%
Huot%
, Martin%
\BCBL {}\ \BBA {} Biondi%
}{%
{\protect \APACyear {2018}}%
}]{%
huot2018automated}
\APACinsertmetastar {%
huot2018automated}%
\begin{APACrefauthors}%
Huot, F.%
, Martin, E\BPBI R.%
\BCBL {}\ \BBA {} Biondi, B.%
\end{APACrefauthors}%
\unskip\
\newblock
\APACrefYearMonthDay{2018}{}{}.
\newblock
{\BBOQ}\APACrefatitle {Automated ambient noise processing applied to fiber
  optic seismic acquisition (DAS)} {Automated ambient noise processing applied
  to fiber optic seismic acquisition (das)}.{\BBCQ}
\newblock
\BIn{} \APACrefbtitle {2018 SEG International Exposition and Annual Meeting.}
  {2018 seg international exposition and annual meeting.}
\PrintBackRefs{\CurrentBib}

\bibitem [\protect \citeauthoryear {%
Karrenbach%
, Cole%
, La~Flame%
\BCBL {}\ \BBA {} Yartsev%
}{%
Karrenbach%
\ \protect \BOthers {.}}{%
{\protect \APACyear {2019}}%
}]{%
karrenbach2019rapid}
\APACinsertmetastar {%
karrenbach2019rapid}%
\begin{APACrefauthors}%
Karrenbach, M.%
, Cole, S.%
, La~Flame, L.%
\BCBL {}\ \BBA {} Yartsev, V.%
\end{APACrefauthors}%
\unskip\
\newblock
\APACrefYearMonthDay{2019}{}{}.
\newblock
{\BBOQ}\APACrefatitle {Rapid Deployment of DAS Fiber-Optic Sensors for
  Monitoring the 2019 Searles Valley and Ridgecrest Earthquakes} {Rapid
  deployment of das fiber-optic sensors for monitoring the 2019 searles valley
  and ridgecrest earthquakes}.{\BBCQ}
\newblock
\BIn{} \APACrefbtitle {AGU Fall Meeting Abstracts} {Agu fall meeting
  abstracts}\ (\BVOL\ 2019, \BPGS\ S31F--0472).
\PrintBackRefs{\CurrentBib}

\bibitem [\protect \citeauthoryear {%
Kingma%
\ \BBA {} Ba%
}{%
Kingma%
\ \BBA {} Ba%
}{%
{\protect \APACyear {2014}}%
}]{%
kingma2014adam}
\APACinsertmetastar {%
kingma2014adam}%
\begin{APACrefauthors}%
Kingma, D\BPBI P.%
\BCBT {}\ \BBA {} Ba, J.%
\end{APACrefauthors}%
\unskip\
\newblock
\APACrefYearMonthDay{2014}{}{}.
\newblock
{\BBOQ}\APACrefatitle {Adam: A method for stochastic optimization} {Adam: A
  method for stochastic optimization}.{\BBCQ}
\newblock
\APACjournalVolNumPages{arXiv preprint arXiv:1412.6980}{}{}{}.
\PrintBackRefs{\CurrentBib}

\bibitem [\protect \citeauthoryear {%
Lay%
\ \protect \BOthers {.}}{%
Lay%
\ \protect \BOthers {.}}{%
{\protect \APACyear {2002}}%
}]{%
lay2002global}
\APACinsertmetastar {%
lay2002global}%
\begin{APACrefauthors}%
Lay, T.%
, Berger, J.%
, Buland, R.%
, Butler, R.%
, Ekstr{\"o}m, G.%
, Hutt, C.%
\BCBL {}\ \BBA {} Romanowicz, B.%
\end{APACrefauthors}%
\unskip\
\newblock
\APACrefYearMonthDay{2002}{}{}.
\newblock
{\BBOQ}\APACrefatitle {Global seismic network design goals update 2002} {Global
  seismic network design goals update 2002}.{\BBCQ}
\newblock
\APACjournalVolNumPages{IRIS, August}{26}{}{}.
\PrintBackRefs{\CurrentBib}

\bibitem [\protect \citeauthoryear {%
Lellouch%
, Lindsey%
, Ellsworth%
\BCBL {}\ \BBA {} Biondi%
}{%
Lellouch%
\ \protect \BOthers {.}}{%
{\protect \APACyear {2020}}%
}]{%
lellouch2020comparison}
\APACinsertmetastar {%
lellouch2020comparison}%
\begin{APACrefauthors}%
Lellouch, A.%
, Lindsey, N\BPBI J.%
, Ellsworth, W\BPBI L.%
\BCBL {}\ \BBA {} Biondi, B\BPBI L.%
\end{APACrefauthors}%
\unskip\
\newblock
\APACrefYearMonthDay{2020}{}{}.
\newblock
{\BBOQ}\APACrefatitle {{Comparison between distributed acoustic sensing and
  geophones: Downhole microseismic monitoring of the FORGE geothermal
  experiment}} {{Comparison between distributed acoustic sensing and geophones:
  Downhole microseismic monitoring of the FORGE geothermal experiment}}.{\BBCQ}
\newblock
\APACjournalVolNumPages{Seismological Research Letters}{91}{6}{3256--3268}.
\newblock
\begin{APACrefDOI} \doi{10.1785/0220200149} \end{APACrefDOI}
\PrintBackRefs{\CurrentBib}

\bibitem [\protect \citeauthoryear {%
Lellouch%
, Schultz%
, Lindsey%
, Biondi%
\BCBL {}\ \BBA {} Ellsworth%
}{%
Lellouch%
\ \protect \BOthers {.}}{%
{\protect \APACyear {2021}}%
}]{%
lellouch2021low}
\APACinsertmetastar {%
lellouch2021low}%
\begin{APACrefauthors}%
Lellouch, A.%
, Schultz, R.%
, Lindsey, N\BPBI J.%
, Biondi, B\BPBI L.%
\BCBL {}\ \BBA {} Ellsworth, W\BPBI L.%
\end{APACrefauthors}%
\unskip\
\newblock
\APACrefYearMonthDay{2021}{}{}.
\newblock
{\BBOQ}\APACrefatitle {Low-magnitude seismicity with a downhole distributed
  acoustic sensing array—Examples from the FORGE geothermal experiment}
  {Low-magnitude seismicity with a downhole distributed acoustic sensing
  array—examples from the forge geothermal experiment}.{\BBCQ}
\newblock
\APACjournalVolNumPages{Journal of Geophysical Research: Solid
  Earth}{126}{1}{e2020JB020462}.
\PrintBackRefs{\CurrentBib}

\bibitem [\protect \citeauthoryear {%
Li%
\ \protect \BOthers {.}}{%
Li%
\ \protect \BOthers {.}}{%
{\protect \APACyear {2021}}%
}]{%
li2021rapid}
\APACinsertmetastar {%
li2021rapid}%
\begin{APACrefauthors}%
Li, Z.%
, Shen, Z.%
, Yang, Y.%
, Williams, E.%
, Wang, X.%
\BCBL {}\ \BBA {} Zhan, Z.%
\end{APACrefauthors}%
\unskip\
\newblock
\APACrefYearMonthDay{2021}{}{}.
\newblock
{\BBOQ}\APACrefatitle {Rapid response to the 2019 Ridgecrest earthquake with
  distributed acoustic sensing} {Rapid response to the 2019 ridgecrest
  earthquake with distributed acoustic sensing}.{\BBCQ}
\newblock
\APACjournalVolNumPages{AGU Advances}{2}{2}{e2021AV000395}.
\PrintBackRefs{\CurrentBib}

\bibitem [\protect \citeauthoryear {%
Lindsey%
\ \protect \BOthers {.}}{%
Lindsey%
\ \protect \BOthers {.}}{%
{\protect \APACyear {2017}}%
}]{%
lindsey2017fiber}
\APACinsertmetastar {%
lindsey2017fiber}%
\begin{APACrefauthors}%
Lindsey, N\BPBI J.%
, Martin, E\BPBI R.%
, Dreger, D\BPBI S.%
, Freifeld, B.%
, Cole, S.%
, James, S\BPBI R.%
\BDBL {}Ajo-Franklin, J\BPBI B.%
\end{APACrefauthors}%
\unskip\
\newblock
\APACrefYearMonthDay{2017}{}{}.
\newblock
{\BBOQ}\APACrefatitle {Fiber-optic network observations of earthquake
  wavefields} {Fiber-optic network observations of earthquake
  wavefields}.{\BBCQ}
\newblock
\APACjournalVolNumPages{Geophysical Research Letters}{44}{23}{11--792}.
\PrintBackRefs{\CurrentBib}

\bibitem [\protect \citeauthoryear {%
Martin%
\ \BBA {} Biondi%
}{%
Martin%
\ \BBA {} Biondi%
}{%
{\protect \APACyear {2018}}%
}]{%
martin2018eighteen}
\APACinsertmetastar {%
martin2018eighteen}%
\begin{APACrefauthors}%
Martin, E\BPBI R.%
\BCBT {}\ \BBA {} Biondi, B\BPBI L.%
\end{APACrefauthors}%
\unskip\
\newblock
\APACrefYearMonthDay{2018}{}{}.
\newblock
{\BBOQ}\APACrefatitle {Eighteen months of continuous near-surface monitoring
  with DAS data collected under Stanford University} {Eighteen months of
  continuous near-surface monitoring with das data collected under stanford
  university}.{\BBCQ}
\newblock
\BIn{} \APACrefbtitle {SEG Technical Program Expanded Abstracts 2018} {Seg
  technical program expanded abstracts 2018}\ (\BPGS\ 4958--4962).
\newblock
\APACaddressPublisher{}{Society of Exploration Geophysicists}.
\PrintBackRefs{\CurrentBib}

\bibitem [\protect \citeauthoryear {%
Martin%
\ \protect \BOthers {.}}{%
Martin%
\ \protect \BOthers {.}}{%
{\protect \APACyear {2017}}%
}]{%
martin2017seismic}
\APACinsertmetastar {%
martin2017seismic}%
\begin{APACrefauthors}%
Martin, E\BPBI R.%
, Castillo, C\BPBI M.%
, Cole, S.%
, Sawasdee, P\BPBI S.%
, Yuan, S.%
, Clapp, R.%
\BDBL {}Biondi, B\BPBI L.%
\end{APACrefauthors}%
\unskip\
\newblock
\APACrefYearMonthDay{2017}{}{}.
\newblock
{\BBOQ}\APACrefatitle {Seismic monitoring leveraging existing telecom
  infrastructure at the SDASA: Active, passive, and ambient-noise analysis}
  {Seismic monitoring leveraging existing telecom infrastructure at the sdasa:
  Active, passive, and ambient-noise analysis}.{\BBCQ}
\newblock
\APACjournalVolNumPages{The Leading Edge}{36}{12}{1025--1031}.
\PrintBackRefs{\CurrentBib}

\bibitem [\protect \citeauthoryear {%
Martin%
\ \protect \BOthers {.}}{%
Martin%
\ \protect \BOthers {.}}{%
{\protect \APACyear {2018}}%
}]{%
martin2018seismic}
\APACinsertmetastar {%
martin2018seismic}%
\begin{APACrefauthors}%
Martin, E\BPBI R.%
, Huot, F.%
, Ma, Y.%
, Cieplicki, R.%
, Cole, S.%
, Karrenbach, M.%
\BCBL {}\ \BBA {} Biondi, B\BPBI L.%
\end{APACrefauthors}%
\unskip\
\newblock
\APACrefYearMonthDay{2018}{}{}.
\newblock
{\BBOQ}\APACrefatitle {A seismic shift in scalable acquisition demands new
  processing: Fiber-optic seismic signal retrieval in urban areas with
  unsupervised learning for coherent noise removal} {A seismic shift in
  scalable acquisition demands new processing: Fiber-optic seismic signal
  retrieval in urban areas with unsupervised learning for coherent noise
  removal}.{\BBCQ}
\newblock
\APACjournalVolNumPages{IEEE Signal Processing Magazine}{35}{2}{31--40}.
\PrintBackRefs{\CurrentBib}

\bibitem [\protect \citeauthoryear {%
Mateeva%
\ \protect \BOthers {.}}{%
Mateeva%
\ \protect \BOthers {.}}{%
{\protect \APACyear {2013}}%
}]{%
mateeva2013distributed}
\APACinsertmetastar {%
mateeva2013distributed}%
\begin{APACrefauthors}%
Mateeva, A.%
, Lopez, J.%
, Mestayer, J.%
, Wills, P.%
, Cox, B.%
, Kiyashchenko, D.%
\BDBL {}Grandi, S.%
\end{APACrefauthors}%
\unskip\
\newblock
\APACrefYearMonthDay{2013}{}{}.
\newblock
{\BBOQ}\APACrefatitle {Distributed acoustic sensing for reservoir monitoring
  with VSP} {Distributed acoustic sensing for reservoir monitoring with
  vsp}.{\BBCQ}
\newblock
\APACjournalVolNumPages{The Leading Edge}{32}{10}{1278--1283}.
\PrintBackRefs{\CurrentBib}

\bibitem [\protect \citeauthoryear {%
NCEDC%
}{%
NCEDC%
}{%
{\protect \APACyear {2014}}%
}]{%
ncedc2014northern}
\APACinsertmetastar {%
ncedc2014northern}%
\begin{APACrefauthors}%
NCEDC.%
\end{APACrefauthors}%
\unskip\
\newblock
\APACrefYearMonthDay{2014}{}{}.
\newblock
{\BBOQ}\APACrefatitle {Northern California earthquake data center} {Northern
  california earthquake data center}.{\BBCQ}
\newblock
\APACjournalVolNumPages{UC Berkeley Seismol Lab Dataset}{}{}{}.
\PrintBackRefs{\CurrentBib}

\bibitem [\protect \citeauthoryear {%
Perol%
, Gharbi%
\BCBL {}\ \BBA {} Denolle%
}{%
Perol%
\ \protect \BOthers {.}}{%
{\protect \APACyear {2018}}%
}]{%
perol2018convolutional}
\APACinsertmetastar {%
perol2018convolutional}%
\begin{APACrefauthors}%
Perol, T.%
, Gharbi, M.%
\BCBL {}\ \BBA {} Denolle, M.%
\end{APACrefauthors}%
\unskip\
\newblock
\APACrefYearMonthDay{2018}{}{}.
\newblock
{\BBOQ}\APACrefatitle {Convolutional neural network for earthquake detection
  and location} {Convolutional neural network for earthquake detection and
  location}.{\BBCQ}
\newblock
\APACjournalVolNumPages{Science Advances}{4}{2}{e1700578}.
\PrintBackRefs{\CurrentBib}

\bibitem [\protect \citeauthoryear {%
Ramachandran%
, Zoph%
\BCBL {}\ \BBA {} Le%
}{%
Ramachandran%
\ \protect \BOthers {.}}{%
{\protect \APACyear {2017}}%
}]{%
ramachandran2017searching}
\APACinsertmetastar {%
ramachandran2017searching}%
\begin{APACrefauthors}%
Ramachandran, P.%
, Zoph, B.%
\BCBL {}\ \BBA {} Le, Q\BPBI V.%
\end{APACrefauthors}%
\unskip\
\newblock
\APACrefYearMonthDay{2017}{}{}.
\newblock
{\BBOQ}\APACrefatitle {Searching for activation functions} {Searching for
  activation functions}.{\BBCQ}
\newblock
\APACjournalVolNumPages{arXiv preprint arXiv:1710.05941}{}{}{}.
\PrintBackRefs{\CurrentBib}

\bibitem [\protect \citeauthoryear {%
Snoek%
, Larochelle%
\BCBL {}\ \BBA {} Adams%
}{%
Snoek%
\ \protect \BOthers {.}}{%
{\protect \APACyear {2012}}%
}]{%
snoek2012practical}
\APACinsertmetastar {%
snoek2012practical}%
\begin{APACrefauthors}%
Snoek, J.%
, Larochelle, H.%
\BCBL {}\ \BBA {} Adams, R\BPBI P.%
\end{APACrefauthors}%
\unskip\
\newblock
\APACrefYearMonthDay{2012}{}{}.
\newblock
{\BBOQ}\APACrefatitle {Practical bayesian optimization of machine learning
  algorithms} {Practical bayesian optimization of machine learning
  algorithms}.{\BBCQ}
\newblock
\BIn{} \APACrefbtitle {Advances in neural information processing systems}
  {Advances in neural information processing systems}\ (\BPGS\ 2951--2959).
\PrintBackRefs{\CurrentBib}

\bibitem [\protect \citeauthoryear {%
Srivastava%
, Hinton%
, Krizhevsky%
, Sutskever%
\BCBL {}\ \BBA {} Salakhutdinov%
}{%
Srivastava%
\ \protect \BOthers {.}}{%
{\protect \APACyear {2014}}%
}]{%
srivastava2014dropout}
\APACinsertmetastar {%
srivastava2014dropout}%
\begin{APACrefauthors}%
Srivastava, N.%
, Hinton, G\BPBI E.%
, Krizhevsky, A.%
, Sutskever, I.%
\BCBL {}\ \BBA {} Salakhutdinov, R.%
\end{APACrefauthors}%
\unskip\
\newblock
\APACrefYearMonthDay{2014}{}{}.
\newblock
{\BBOQ}\APACrefatitle {Dropout: a simple way to prevent neural networks from
  overfitting.} {Dropout: a simple way to prevent neural networks from
  overfitting.}{\BBCQ}
\newblock
\APACjournalVolNumPages{Journal of Machine Learning
  Research}{15}{1}{1929--1958}.
\PrintBackRefs{\CurrentBib}

\bibitem [\protect \citeauthoryear {%
Stoffer%
}{%
Stoffer%
}{%
{\protect \APACyear {2006}}%
}]{%
stoffer2006s}
\APACinsertmetastar {%
stoffer2006s}%
\begin{APACrefauthors}%
Stoffer, P\BPBI W.%
\end{APACrefauthors}%
\unskip\
\newblock
\APACrefYear{2006}.
\newblock
\APACrefbtitle {Where's the San Andreas Fault?: A Guidebook to Tracing the
  Fault on Public Lands in the San Francisco Bay Region} {Where's the san
  andreas fault?: A guidebook to tracing the fault on public lands in the san
  francisco bay region}\ (\BVOL~16).
\newblock
\APACaddressPublisher{}{US Geological Survey}.
\PrintBackRefs{\CurrentBib}

\bibitem [\protect \citeauthoryear {%
{The GPyOpt authors}%
}{%
{The GPyOpt authors}%
}{%
{\protect \APACyear {2016}}%
}]{%
gpyopt2016}
\APACinsertmetastar {%
gpyopt2016}%
\begin{APACrefauthors}%
{The GPyOpt authors}.%
\end{APACrefauthors}%
\unskip\
\newblock
\APACrefYearMonthDay{2016}{}{}.
\newblock
\APACrefbtitle {{GPyOpt}: A Bayesian Optimization framework in python.}
  {{GPyOpt}: A bayesian optimization framework in python.}
\newblock
\APAChowpublished {\url{http://github.com/SheffieldML/GPyOpt}}.
\PrintBackRefs{\CurrentBib}

\bibitem [\protect \citeauthoryear {%
Zhang%
\ \protect \BOthers {.}}{%
Zhang%
\ \protect \BOthers {.}}{%
{\protect \APACyear {2019}}%
}]{%
zhang2019aftershock}
\APACinsertmetastar {%
zhang2019aftershock}%
\begin{APACrefauthors}%
Zhang, Q.%
, Xu, T.%
, Zhu, H.%
, Zhang, L.%
, Xiong, H.%
, Chen, E.%
\BCBL {}\ \BBA {} Liu, Q.%
\end{APACrefauthors}%
\unskip\
\newblock
\APACrefYearMonthDay{2019}{}{}.
\newblock
{\BBOQ}\APACrefatitle {Aftershock detection with multi-scale description based
  neural network} {Aftershock detection with multi-scale description based
  neural network}.{\BBCQ}
\newblock
\BIn{} \APACrefbtitle {2019 IEEE International Conference on Data Mining
  (ICDM)} {2019 ieee international conference on data mining (icdm)}\ (\BPGS\
  886--895).
\PrintBackRefs{\CurrentBib}

\bibitem [\protect \citeauthoryear {%
Zheng%
, Harris%
, Li%
\BCBL {}\ \BBA {} Al-Rumaih%
}{%
Zheng%
\ \protect \BOthers {.}}{%
{\protect \APACyear {2020}}%
}]{%
zheng2020sc}
\APACinsertmetastar {%
zheng2020sc}%
\begin{APACrefauthors}%
Zheng, J.%
, Harris, J\BPBI M.%
, Li, D.%
\BCBL {}\ \BBA {} Al-Rumaih, B.%
\end{APACrefauthors}%
\unskip\
\newblock
\APACrefYearMonthDay{2020}{}{}.
\newblock
{\BBOQ}\APACrefatitle {SC-PSNET: A deep neural network for automatic P-and
  S-phase detection and arrival-time picker using 1C recordings} {Sc-psnet: A
  deep neural network for automatic p-and s-phase detection and arrival-time
  picker using 1c recordings}.{\BBCQ}
\newblock
\APACjournalVolNumPages{Geophysics}{85}{4}{U87--U98}.
\PrintBackRefs{\CurrentBib}

\end{thebibliography}
%
% don't specify bibliographystyle

% In the References section, cite the data/software described in the Availability Statement (this includes primary and processed data used for your research). For details on data/software citation as well as examples, see the Data & Software Citation section of the Data & Software for Authors guidance
% https://www.agu.org/Publish-with-AGU/Publish/Author-Resources/Data-and-Software-for-Authors#citation

%%%%%%%%%%%%%%%%%%%%%%%%%%%%%%%%%%%%%%%%%%%%%%%

%\bibliography{enter your bibtex bibliography filename here}

%Reference citation instructions and examples:
%
% Please use ONLY \cite and \citeA for reference citations.
% \cite for parenthetical references
% ...as shown in recent studies (Simpson et al., 2019)
% \citeA for in-text citations
% ...Simpson et al. (2019) have shown...
%
%
%...as shown by \citeA{jskilby}.
%...as shown by \citeA{lewin76}, \citeA{carson86}, \citeA{bartoldy02}, and \citeA{rinaldi03}.
%...has been shown \cite{jskilbye}.
%...has been shown \cite{lewin76,carson86,bartoldy02,rinaldi03}.
%... \cite <i.e.>[]{lewin76,carson86,bartoldy02,rinaldi03}.
%...has been shown by \cite <e.g.,>[and others]{lewin76}.
%
% apacite uses < > for prenotes and [ ] for postnotes
% DO NOT use other cite commands (e.g., \citet, \citep, \citeyear, \citealp, etc.).
% \nocite is okay to use to add references from your Supporting Information
%

\end{document}